\documentclass[%
 reprint,
 amsmath,amssymb,longbibliography,
 aps,
prl,
floatfix,
]{revtex4-2}
\usepackage{graphicx}
\usepackage{bm}
\usepackage{comment}

\newcommand{\chir}{\chi(r)}

\usepackage{xcolor}

\begin{document}

\title{Density depletion and enhanced fluctuations in water near hydrophobic solutes: identifying the underlying physics}%
\author{Mary K. Coe}
\author{Robert Evans}%
 \author{Nigel B. Wilding}%
 \email[Corresponding author: ]{nigel.wilding@bristol.ac.uk}

\affiliation{%
 H.H. Wills Physics Lab, University of Bristol, Tyndall Avenue, Bristol BS8 1TL. U.K.}

\begin{abstract}

We investigate the origin of the density depletion and enhanced density fluctuations that occur in water in the vicinity of an extended hydrophobic solute. We argue that both phenomena are remnants of the critical drying surface phase transition that occurs at liquid-vapor coexistence in the macroscopic planar limit, ie.~as the solute radius $R_s\to\infty$. Focusing on the density profile $\rho(r)$ and a sensitive spatial measure of fluctuations, the local compressibility profile $\chi(r)$, we develop a scaling theory which expresses the extent of the density depletion and enhancement in compressibility in terms of $R_s$, the strength of solute-water attraction $\varepsilon_s$, and the deviation from liquid-vapor coexistence $\delta\mu$. Testing the predictions  against results of classical density functional theory for a simple solvent and Grand Canonical Monte Carlo simulations of a popular water model, we find that the theory provides a firm physical basis for understanding how water behaves at a hydrophobe. 

\end{abstract}

\maketitle


The term hydrophobicity is used in a variety of different contexts to describe phenomena which are driven by an aversion to water.  Typically the problems of interest are delineated by their characteristic length scales  \cite{Chandler:2005aa}. For instance, when considering the properties of a planar solid substrate, the degree of hydrophobicity is quantified by the macroscopic contact angle $\theta$ that a sessile water drop makes with the surface: the extreme hydrophobic limit corresponds to $\theta\to 180^\circ$ \cite{Simpson_2015}. In soft matter settings, the hydrophobic effect refers to the thermodynamic tendency of amphiphilic molecules in solution to self assemble, while in biology the hydrophobicity of proteins in an aqueous environment is believed to play a central role in their folding behaviour \cite{LevyOnuchic2006, Jamadagni2011}.  Hydrophobicity is crucially important for understanding solvation properties in physio-chemical systems,  eg.~when hydrophobic molecular solutes of various sizes are immersed in water  \cite{Huang:2002xr,Lum:1999aa,HuangChandler2000PNAS,HuangGeisslerChandler2001,Grdadolnik,Jeanmairet2015}. Beyond solubility, one is interested in the perturbations to the water structure that occur in the solute's vicinity, see  eg.~\cite{Grdadolnik,SongMolinero2013,SchnupfBrady2017,HuangBerne2003}.

For water near a macroscopic hydrophobic planar substrate, experiments report a region of depleted density \cite{Mezger:2006zl, Ocko:2008fv, Mezger:2010lq} whose extent continues to be controversial \cite{Chattopadhyay:2010aa}. Recent studies have revealed new insight: the extreme hydrophobic limit $\theta=180^\circ$ corresponds to a critical surface phase transition known as `drying'~\cite{EvansStewartWilding2016,EvansStewartWilding2017,EvansStewartWilding2019}.  When the substrate-fluid attraction is sufficiently small to yield the near limiting contact angle, water (or indeed any solvent) develops a vapor (or drying) layer near the substrate surface as the system approaches bulk vapor-liquid coexistence.   Precisely at the drying point, the equilibrium thickness of the vapor layer $\ell_{eq}$ diverges. This can be quantified by a profile $\rho(z)$ which measures the local number density at a distance $z$ from the (planar) substrate. However, at standard temperature and pressure STP (ambient) conditions, water is not quite at coexistence; it has a very small but non-zero supersaturation: the chemical potential deviation $\delta \mu$ from coexistence $\mu_{co}$ is $\beta\delta\mu = \beta(\mu-\mu_{co})\approx10^{-3},\; \beta=1/k_BT$~\cite{cerdeirina2011}. Accordingly a drying layer of only finite thickness can form at a strongly hydrophobic substrate. A further key feature of a critical drying transition is the occurrence of pronounced density fluctuations. These are quantified rigorously by the local compressibility profile $\chi(z)=\partial \rho(z)/\partial \mu|_T$, which displays a maximum that diverges on the approach to the drying point and which, in common with the divergence of $\ell_{eq}$, exhibits critical scaling behaviour~\cite{EvansStewart2015,EvansStewartWilding2017,EvansStewartWilding2016}. The nature of the scaling behaviour, including the values of associated critical exponents, has recently been found (in the planar case)  to be sensitive to the range of the fluid-fluid and substrate-fluid interparticle forces~\cite{EvansStewartWilding2019}. 

\begin{figure}[h]
\includegraphics[width=0.51\textwidth]{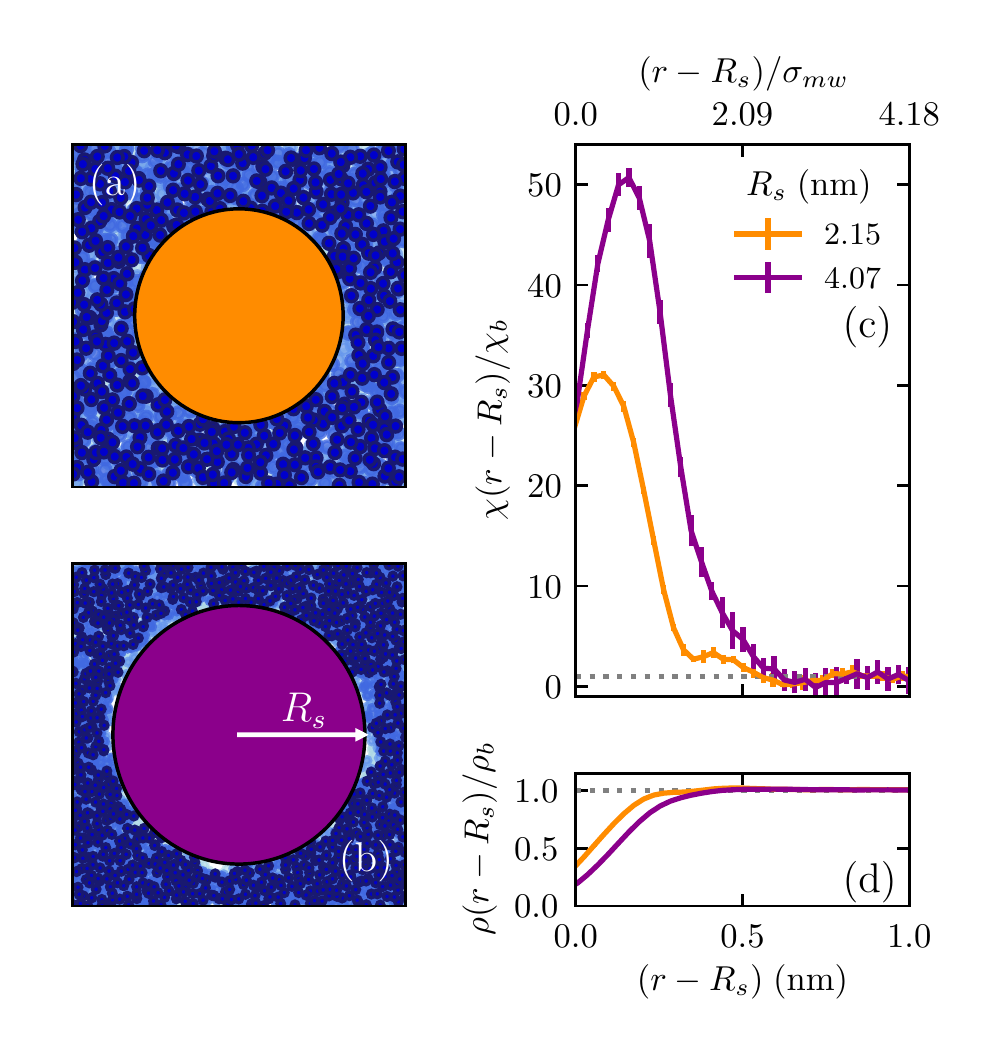}
\caption{Cross-sections of a snapshot of  mw water molecules (diameter $\sigma_{mw}=0.239$\:nm) surrounding a hard spherical solute of radius  $R_s=2.15$\:nm {\bf (a)}, and   $R_s=4.07$\:nm {\bf (b)}. {\color{black}The shade of the mw water molecules lightens the greater their distance from the cross-sectional plane.} The corresponding  average local compressibility profiles $\chir$ {\bf (c)} and  density profiles $\rho(r)$ {\bf (d)}, scaled by their bulk values. The profile line color matches that of the corresponding solute. Temperature $T=426$K, $\beta\delta\mu=10^{-3}$. }
\label{fig:solute_profile_comparison}
\end{figure}

For water near a finite hydrophobic solute, simulations indicate that when the solute is extended, i.e. the ratio of solute diameter to water molecule diameter $\sigma_s/\sigma_w \gtrsim 3$, a region of depleted density develops together with `enhanced' density fluctuations. Representative papers are Refs.~\cite{Lum:1999aa,Huang:2000wq,Sarupria2009,AcharyaGarde2010,PatelVarillyChandler2010,Mittal:2008aa,Oleinikova:2012aa,VaikuntanathanE2224}.  Fig.~\ref{fig:solute_profile_comparison}, which depicts our Grand Canonical Monte Carlo (GCMC) simulation measurements of $\rho(r)$ and $\chir$ for a monatomic water model (see below) in contact with a hard spherical solute for two different radii $R_s$,  exposes clearly  and quantifies  much more accurately than previous treatments the density depletion and substantial enhancement of density fluctuations. Increasing $R_s$ amplifies both effects. The accompanying snapshots reveal that the growth in the extent of the density depletion and in the maximum in $\chir$ are associated with the appearance of extended vapor bubbles, strongly fluctuating during the simulation, on the solute surface.  Such effects for finite hydrophobes are reminiscent of those occurring at a macroscopic planar substrate near critical drying~\cite{EvansStewartWilding2017}. (Note also that there is experimental evidence for nanobubbles at hydrophobic surfaces: \cite{Tyrrell:2001aa,SteitzFindenegg2003}).  To date, however, there is no unifying and comprehensive theoretical framework  linking the variety of hydrophobic phenomena that occur across the disparate length scales extending from nanoscale hydrophobes to planar substrates. Specifically, existing theories eg.~\cite{Lum:1999aa,PatelGarde2012,VaikuntanathanE2224} take no account of the existence and central role of the critical drying transition that occurs in the limit of a planar substrate with vanishing,  or very weak substrate-fluid attraction \cite{EvansStewartWilding2016,EvansStewartWilding2017, EvansStewartWilding2019} \footnote{Mention is made in the literature, e.g. ~\cite{Lum:1999aa,PatelGarde2012,VaikuntanathanE2224} of `drying' or `dewetting' in relation to the observed phenomena, but these terms appear to be used loosely as a shorthand for the appearance of a region of depleted density, particularly one spanning the region between hydrophobic solute. {\color{black} The commentary by R. Evans, Faraday Discussions, {\bf{146}}, 297 (2010) emphasizes why nomenclature  is important. This will be discussed further in a future publication}}.

Using binding potential arguments similar to those of Refs.~\cite{EvansStewartWilding2017,EvansStewartWilding2019}, we develop a scaling theory that describes how the extent of the density depletion $\ell_{eq}$ and the enhancement of fluctuations depend on the three quantities that together control the proximity to the drying point: namely the supersaturation $\delta\mu$, the strength of solute-water attractions $\varepsilon_s$, and the curvature $R_s^{-1}$ of the solute. By encapsulating the curvature as an effective scaling field we forge the link between the macroscopic (planar substrate) limit and the microscopic solute case. 

We test our predictions by implementing: i) extensive classical density functional theory (DFT) calculations for a solvophobic spherical solute of radius $R_s$ immersed in a simple (truncated) Lennard-Jones (LJ) liquid and  ii)  GCMC simulations of a solute-in-water system using the popular `monatomic water' (mw) model \cite{Molinero:2009aa}. In both systems the thermodynamic conditions are comparable to ambient water. Our DFT and simulation results confirm the scaling predictions, and show that the fluctuations occurring in water near an extended hydrophobic solute are a remnant of the density fluctuations, characterized by a diverging parallel correlation length, that occur at critical drying for a planar substrate.  

Our theory follows the treatment of drying at a (very large) hard-sphere in \cite{Evans:2004aa,StewartPRE2005},  but we include crucial solute-fluid attractions and analyze fluctuations using the local compressibility.  We consider $\omega_{ex}(l|{R_s})$, the excess grand potential per unit area of the solute-fluid interface that surrounds a solute of radius $R_s$, as a function of the thickness $\ell$ of the intruding vapor layer. For a liquid  governed by short-ranged interparticle forces, e.g. a truncated LJ model or mw, and long-ranged solute-liquid interactions, we expect \cite{SM2021}{\color{black} (for background see \cite{Dietrich1988,Schick1988,BiekerDietrich1998})}

\begin{equation}
    \omega_{ex}(\ell|R_s) = \gamma_{sv} + \gamma_{lv} + a e^{-\ell/\xi} + \frac{b}{\ell^2} + \tilde{p}\ell 
    \label{bp_SR_ff_LR_wf_spherical}
\end{equation}
where 
\begin{equation}
  \tilde{p} = \delta\mu\delta\rho + \frac{2\gamma_{lv}}{R_s},
\end{equation}
with $\delta\rho=\rho_l-\rho_v$ the difference between the liquid and vapor
densities at coexistence. $\gamma_{sv}$ and $\gamma_{lv}$ are the planar substrate-vapor and liquid-vapor surface tensions, respectively. The exponential term accounts for short-ranged
fluid-fluid interactions; $\xi$ is the correlation length of the bulk vapor, and $a$ is a positive coefficient. The term $b/l^2$ is associated with the long ranged van der Waals (dispersion) forces existing between the solute and the fluid. The constant $b<0$ is proportional to  $\varepsilon_s$, as described in the Supplementary Material (SM) \cite{SM2021} {\color{black} which also includes refs.~\cite{Israelachvili,ChackoEvansArcher2017,Evans1979,Roth:2010vn,EvansFundInhomFluids,StewartThesis,StewartEvans2005,Wilding1995,XuMolinero2010,RussoAkahaneTanaka2018}}. $\tilde{p}$ has a natural interpretation: it is a pressure that combines the supersaturation with the Laplace pressure from the presence of (an incipient) liquid-vapor interface. The final term in (\ref{bp_SR_ff_LR_wf_spherical}) occurs generally; it does not depend on the choice of potentials. Minimising $\omega_{ex}(\ell|R_s)$ with respect to $\ell$ yields an equation for the equilibrium vapor layer thickness

\begin{equation}
    -\frac{\ell_{eq}(R_s)}{\xi} = \ln \left(\frac{\xi}{a}\right) + \ln\left(\tilde{p} - \frac{2b}{[\ell_{eq}(R_s)]^3}\right)\:.
    \label{bp_SR_ff_LR_wf_leq_spherical}
\end{equation}
We now omit the explicit dependence of $\ell_{eq}$ on $R_s$. The maximum in the local compressibility $\chir$ occurs in the proximity of $r=\ell_{eq}$ and is given by \cite{SM2021}
\begin{equation}
    \chi(\ell_{eq}|R_s) =\left.\frac{\partial\rho(r)}{\partial \mu}\right|_{T,l=\ell_{eq}}= -\rho^\prime(R_s+\ell_{eq})\left.\frac{\partial \ell_{eq}}{\partial\mu}\right|_T
\end{equation}
where the prime denotes differentiation with respect to $r$. Evaluating the derivative w.r.t. $\mu$ yields
\begin{equation}
    \chi(\ell_{eq}|R_s) = \xi\delta\rho\rho'(R_s+\ell_{eq})\left(\tilde{p} - \frac{2b}{\ell_{eq}^3}\left(1-\frac{3\xi}{\ell_{eq}}\right)\right)^{-1}\:.
    \label{bp_SR_ff_LR_wf_chi_mu_spherical}
\end{equation}
Critical drying requires a planar substrate ($R_s^{-1}=0$) and liquid-vapor coexistence ($\delta\mu= 0$). Given both conditions, one finds that as $b\to 0$, i.e.~the substrate-fluid attraction $\varepsilon_s$ vanishes and the substrate reduces to a hard wall, $\ell_{eq}$ diverges as $\ell_{eq}/\xi=-\ln \varepsilon_s+3\ln (\ell_{eq}/\xi)$, which implies a value of the (adsorption) critical exponent $\beta_s=0$. The maximum of the local compressibility is predicted to diverge like $\chi(\ell_{eq}) \sim\varepsilon_s^{-1}$. Other critical exponents are laid out in~\cite{EvansStewartWilding2017}. All divergences are absent if either of $-b$ (proportional to $\varepsilon_s$ ) or $\delta\mu$ are non-vanishing.

For a solute of finite radius, Eqs.~(\ref{bp_SR_ff_LR_wf_leq_spherical}) and (\ref{bp_SR_ff_LR_wf_chi_mu_spherical}) show that the dominant effect of the curvature, $R_s^{-1}$, is to introduce an additional effective pressure which shifts the system away from the critical drying point. Accordingly $R_s^{-1}$ acts with the supersaturation $\delta\mu$ to determine the value of the scaling field $\tilde{p}$, which together with the value of $\varepsilon_s$, hence $b$, controls deviations from criticality. It follows that provided the magnitudes of $-b$, $R_s^{-1}$ and $\delta\mu$ are all sufficiently small, the system is expected to display {\em near-critical} behavior. The range of values of $R_s$ and $\varepsilon_s$ for which such behaviour arises, follows from Eq.~(\ref{bp_SR_ff_LR_wf_chi_mu_spherical}): if $2\gamma_{lv}/R_s\ll 2b/\ell_{eq}^3$, then the local compressibility at coexistence grows as $\chi(\ell_{eq}| R_s,\delta\mu=0)\sim \varepsilon_s^{-1}$, the growth expected for a planar substrate. However, if $2\gamma_{lv}/R_s \gg 2b/\ell_{eq}^3$, then $\chi(\ell_{eq}|R_s,\delta\mu=0)\sim R_s$; the curvature constrains dramatically the growth of density fluctuations. 

Eqs.~(\ref{bp_SR_ff_LR_wf_leq_spherical}) and (\ref{bp_SR_ff_LR_wf_chi_mu_spherical}) provide explicit predictions for the near-critical scaling of $\ell_{eq}$ and $\chi(\ell_{eq}|R_s)$ with respect to $\tilde{p}$ and  $b$ (equivalent to $\varepsilon_s$). We gain quantitative insight into the range of parameters for which scaling behaviour occurs, by performing DFT calculations for a simple model fluid and GCMC simulations of a water model in contact with a spherical solute. 

\begin{figure}[htb]
    \centering
    \includegraphics[width=0.48\textwidth]{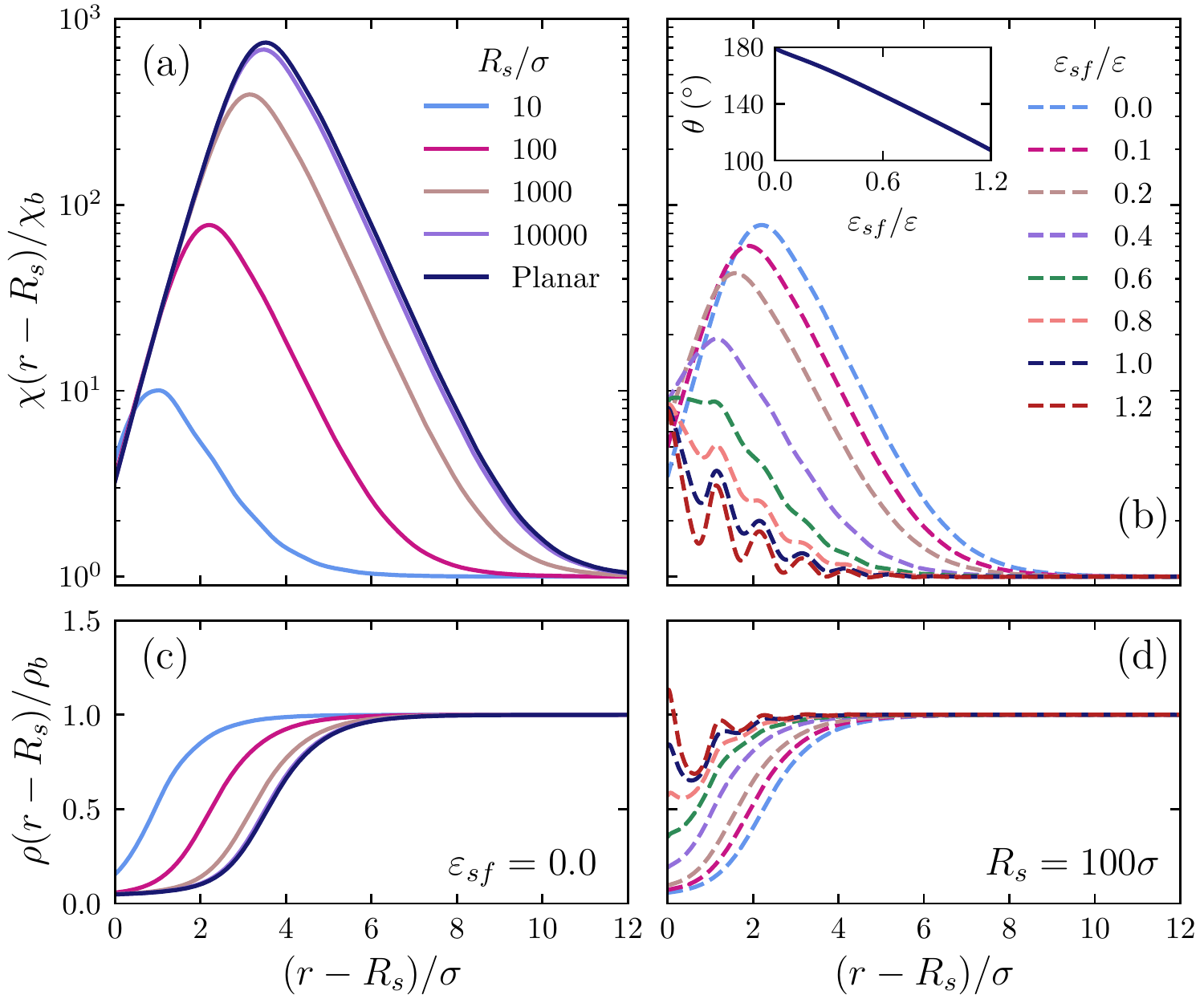}
    \caption{DFT results for {\bf (a,b)} $\chi(r)$, and {\bf (c,d)} $\rho(r)$, scaled by their bulk values, for a truncated LJ model liquid in contact with a spherical solute. Panels {\bf (a,c)} show the effect of varying solute radius $R_s$ for constant $\varepsilon_{sf}=0.0$ - the  hard sphere solute. Panels {\bf (b,d)} show the effect of increasing attraction $\varepsilon_{sf}$ for constant $R_s=100\sigma$. In all cases $T=0.775T_c$ and $\beta\delta\mu=10^{-3}$. The inset shows how the contact angle $\theta$, pertaining to bulk coexistence and the planar limit, depends on $\varepsilon_{sf}$.}

    \label{fig:DFTprofiles}
\end{figure}

Our DFT calculations parallel those for infinite planar geometry~\cite{EvansStewartWilding2017}, modified to treat a spherical solute centred at the origin as described in the SM \cite{SM2021}. DFT enables calculation of the profiles $\rho(r)$ and $\chi(r)$ for a wide range of $R_s$, $\varepsilon_s$ and supersaturations. Results are presented for a truncated LJ model liquid with well-depth $\varepsilon$ and diameter $\sigma$ at temperature $T=0.775T_c$, where $T_c$ is the bulk critical temperature, and  supersaturation $\beta\delta\mu=10^{-3}$, similar to the value for water at STP. Note that results are presented in terms of the ratio $\varepsilon_{sf}/\varepsilon$ where the integrated solute-fluid attraction strength $\varepsilon_{sf}$ is proportional to the solute particle-fluid particle attractive well depth $\varepsilon_s$ - see Eq. (11) of the SM \cite{SM2021}. Fig.~\ref{fig:DFTprofiles}(a,c) plot the variations of the profiles with $R_s$ for a hard spherical solute ($\varepsilon_s=0$), while fig.~\ref{fig:DFTprofiles}(b,d) plot the dependence on solute-fluid attraction ($\varepsilon_{sf}/\varepsilon)$ for a fixed solute radius $R_s=100\sigma$. The results clearly display near-critical behaviour. Fig.~\ref{fig:DFTprofiles}(c) reflects extended density depletion around the solute as $R_s$ is increased (since $\delta\mu>0$ the thickness $\ell_{eq}$ cannot diverge even in the planar limit). Similarly the local compressibility shown in Fig.~\ref{fig:DFTprofiles}(a) exhibits a peak growing to orders of magnitude greater than the bulk value as $R_s\to\infty$. As the radius of the solute is increased, the profiles tend smoothly to those  at  a  planar  wall, confirming that  solute curvature is a natural thermodynamic measure of deviation from critical drying.  The behaviour as a function of $\varepsilon_{sf}$ (Fig.~\ref{fig:DFTprofiles}(b,d))  also points to near-critical behavior: reducing $\varepsilon_{sf}(\propto\varepsilon_s)$ increases the degree of hydrophobicity leading to a more extended depleted density region and more enhanced density fluctuations. For stronger attraction, i.e.  for $\varepsilon_{sf}/\varepsilon > 0.6$, pronounced oscillations develop in $\chi(r)$ reflecting those in $\rho(r)$ and there is no longer a distinct maximum signalling solvo/hydrophobic behaviour.  Notwithstanding,  for all values of $\varepsilon_{sf}/\varepsilon$ presented in Fig. (\ref{fig:DFTprofiles}b) the profiles display behaviour distinct from those of solvo/hydrophilic systems. This implies that the influence of critical drying extends much further than just the critical scaling region. Contour plots, shown in the SM \cite{SM2021}, elucidate further how $\ell_{eq}$ and $\chi(\ell_{eq}|R_s)$ depend on $R_s$ and $\varepsilon_s$.

\begin{figure}
    \centering
    \includegraphics[width = 0.49
    \textwidth]{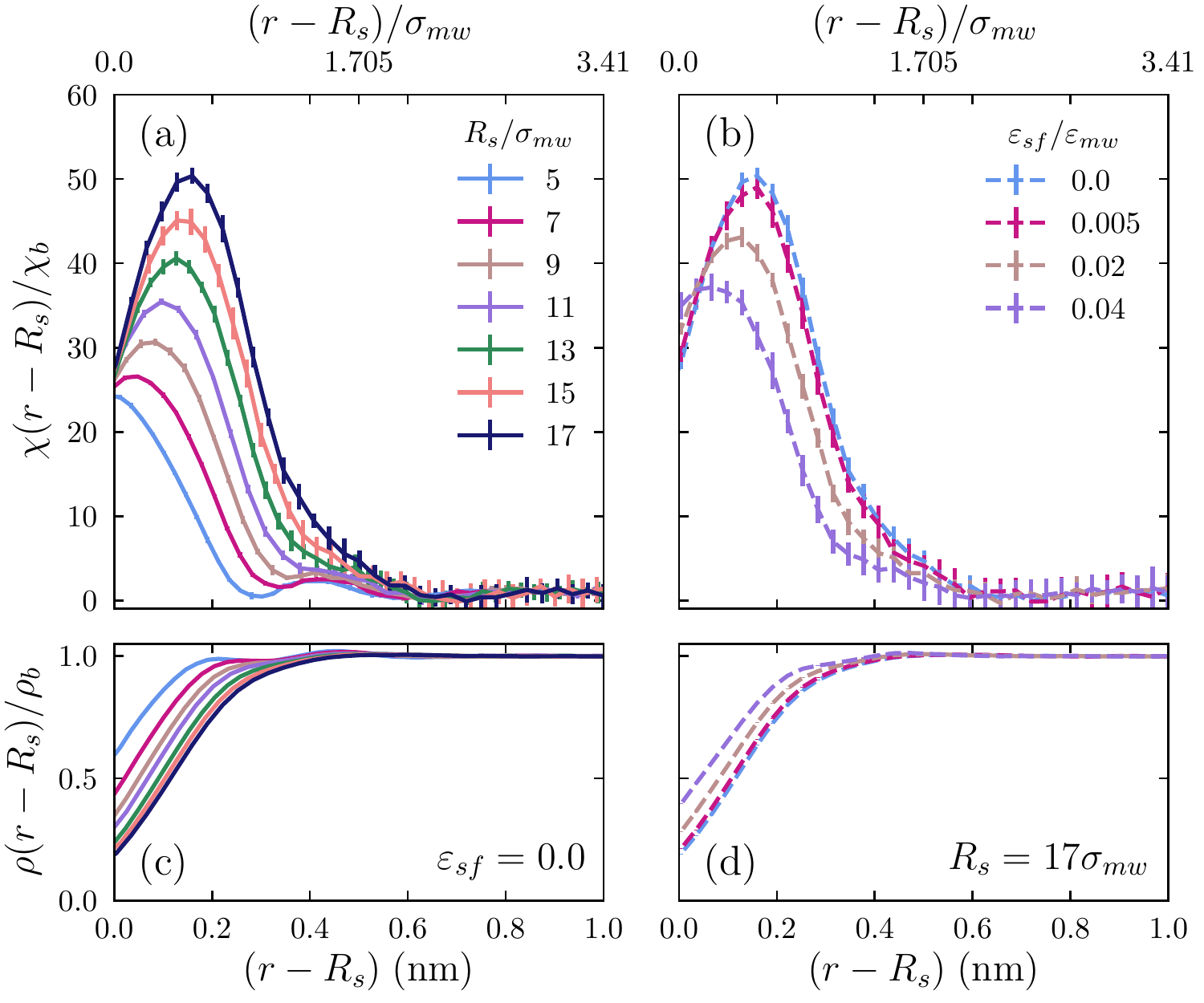}
    \caption{GCMC results for {\bf (a,b)}  $\chi(r)$, and {\bf (c,d)} $\rho(r)$, scaled by their bulk values, for monatomic water (diameter $\sigma_{mw} =0.239$ nm) in contact with spherical solutes. Panels (a,c)  refer to a hard solute for $7$ solute radii $R_s$, spanning the interval $(1.2:4.06)$\:nm. Panels (b,d) refer to  four values of the substrate-fluid attraction $\varepsilon_{sf}$ and constant radius $R_s=17\sigma_{mw}$.}
       \label{fig:mW_profiles}
\end{figure}

  GCMC simulation studies were carried out for the monatomic water (mw) model \cite{Molinero:2009aa}  described in the SM \cite{SM2021}. An accurate determination of the bulk vapor-liquid line and critical point of the model in the $\mu-T$ plane, was performed using histogram methods following Ref.~\cite{Wilding1995,SM2021}. For the study of hydrophobic solutes, we imposed thermodynamic conditions appropriate to real water at STP, setting $\beta\delta\mu=10^{-3}$ and $T/T_c=0.464$ \footnote{An accurate determination of the coexistence line for a particular water model is vital to prescribe the small supersaturation pertinent in ambient water which governs the extent of the density depletion and magnitude of enhanced density fluctuations associated with near-critical drying.
  }. Fig.~\ref{fig:mW_profiles} shows $\rho(r)$ and $\chi(r)$ measured for solutes with various $R_s$ and integrated solute-water attraction $\varepsilon_{sf}$. Note we use the same form of solute-fluid attractions as in the case of DFT, given in Eq. (11) of the SM \cite{SM2021}. For extended solutes having $R_s\gtrsim 1$\:nm,  a region of depleted density region emerges accompanied by enhanced density fluctuations reflected in a pronounced peak in $\chir$ which grows in height and moves further away from the solute as $R_s$ increases, mirroring the trend observed in the DFT calculations (Fig.~\ref{fig:DFTprofiles}). Increasing $\varepsilon_{sf}$ reduces the degree of hydrophobicity, decreasing both the extent of the density depletion and the magnitude of the local compressibility maximum. Although the range of solute radii accessible to simulations is much smaller than for DFT, one sees that for a given solute to water size ratio, the extent of the density depletion and the maximum in $\chir$ are comparable with the DFT results, implying that the trends, and therefore the physics,  found in our  water model simulations mirror those found in the DFT calculations for a much simpler model system.

\begin{figure}
    \centering
    \includegraphics[width=0.5\textwidth]{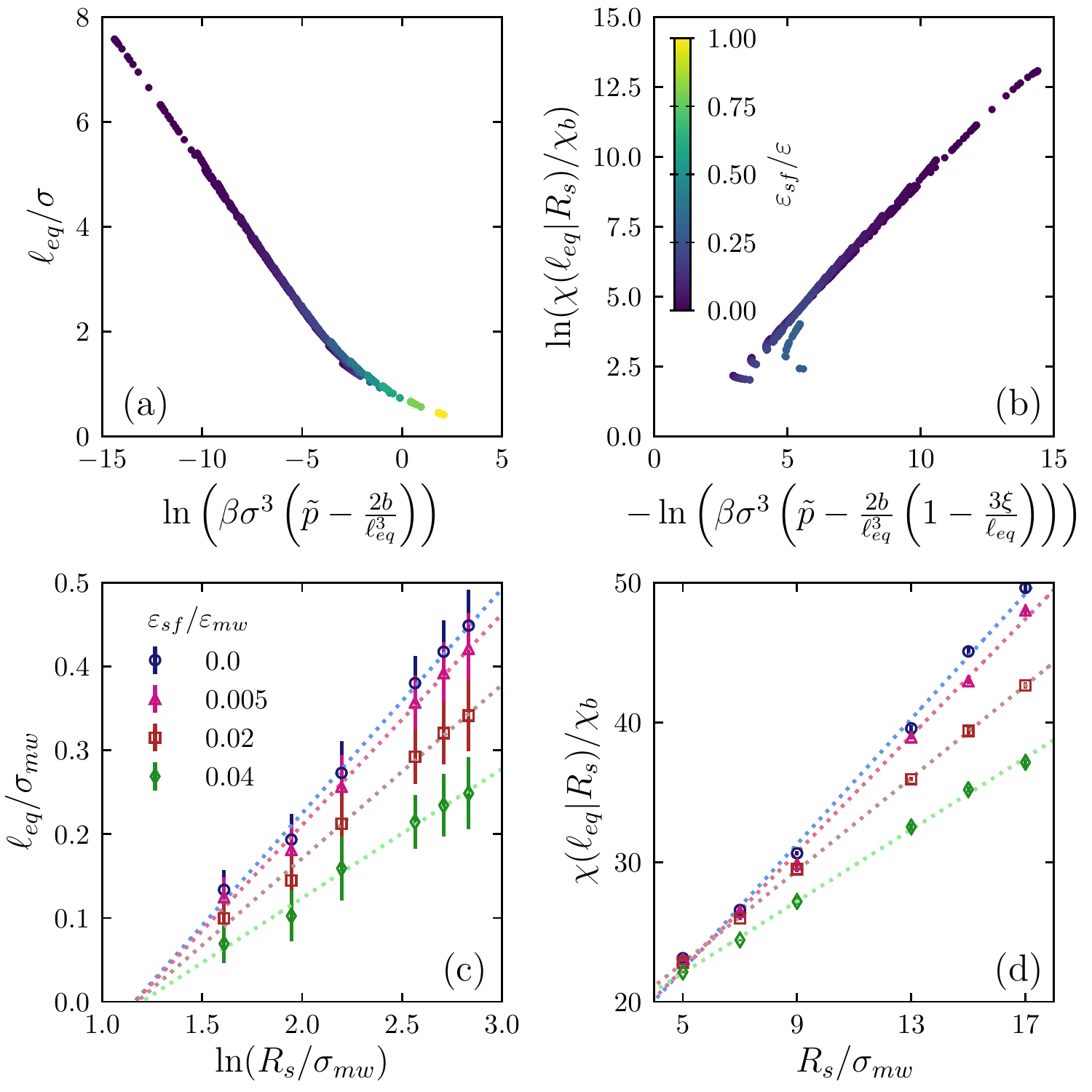}
    \caption{{\bf (a,b)} DFT results for values of $\beta\delta\mu$ and $R_s$ spanning the intervals $(10^{-6},10^{-3})$ and $(10\sigma,\infty)$ respectively, and various $\varepsilon_{sf}/\varepsilon$ as shown on the color chart, testing scaling predictions
    (\ref{bp_SR_ff_LR_wf_leq_spherical}) and (\ref{bp_SR_ff_LR_wf_chi_mu_spherical}). {\bf(c,d)} GCMC results for mw at $\beta\delta\mu=10^{-3},T/T_c=0.464$ testing the scaling predictions in the computationally accessible curvature limit. The values of $\varepsilon_{sf}/\varepsilon_{mw}$ are given in the legend. }
    \label{fig:scaling}
\end{figure}

We have tested the scaling predictions for $\ell_{eq}$ and $\chi(\ell_{eq}|R_s)$, i.e. Eqs.~(\ref{bp_SR_ff_LR_wf_leq_spherical}) and (\ref{bp_SR_ff_LR_wf_chi_mu_spherical}), using both our DFT and simulation data. For $\ell_{eq}$, the linear behaviour shown in the main panel of Fig.~(\ref{fig:scaling}a) confirms that within DFT, (\ref{bp_SR_ff_LR_wf_leq_spherical}) is obeyed for sufficiently small values of $\beta\delta\mu,\: \varepsilon_{sf}$, and $R_s^{-1}$. {\color{black}Remarkably, these predictions also appear to be obeyed down to microscopic values of $\ell_{eq}$. Our mw simulations cannot access very large $R_s$, and in the computationally accessible regime the scaling should be dominated by the Laplace ($R_s^{-1}$) term entering $\tilde{p}$ in (\ref{bp_SR_ff_LR_wf_leq_spherical}) and (\ref{bp_SR_ff_LR_wf_chi_mu_spherical}). Fig.~(\ref{fig:scaling}c) shows that $\ell_{eq}$ grows as $\ln(R_s)$ as $\varepsilon_{sf}\to 0$, in accordance with (\ref{bp_SR_ff_LR_wf_leq_spherical}). Corresponding results for $\chi(\ell_{eq}|R_s)$ in Fig.~(\ref{fig:scaling}b,d) confirm the scaling predictions of (\ref{bp_SR_ff_LR_wf_chi_mu_spherical})}.

In summary, we have argued that the physics that underlies density depletion and enhanced fluctuations in water at extended hydrophobic solutes is linked intimately to the critical drying transition that occurs for a general, strongly solvophobic substrate in the planar limit at vapor-liquid coexistence. The influence of critical drying extends over a range of off-coexistence state points and to solutes of finite radius, leading to near-critical scaling behaviour, the nature of which we have clarified. {\color{black}We have identified how, for strongly hydrophobic solutes, phenomena at microscopic length scales depend on the solute size and the strength of solute-water attraction.}

\begin{acknowledgments}
This work used the facilities of the Advanced Computing Research Centre, University of Bristol. We thank F. Turci for valuable discussions. R.E. acknowledges Leverhulme Trust Grant EM-2020-029\textbackslash 4.
\end{acknowledgments}

\bibliographystyle{apsrev4-1}
\providecommand{\noopsort}[1]{}\providecommand{\singleletter}[1]{#1}%

\end{document}